\documentclass[sigconf]{acmart}

\renewcommand\footnotetextcopyrightpermission[1]{} 
\title{Designing for Being-With: Presence Without Personhood in Conversational Human--AI Interaction}

\author{Hector Michael Fried}
\affiliation{%
  \institution{University of Edinburgh}
  \city{Edinburgh}
  \country{United Kingdom}
}

\author{Robin Hill}
\affiliation{%
  \institution{University of Edinburgh}
  \city{Edinburgh}
  \country{United Kingdom}
}

\begin{abstract}
Conversational AI systems increasingly generate social presence through linguistic fluency, emotional mirroring, and continuity across interactions. While these qualities can support engagement, they also risk relational overreach---particularly in care-adjacent contexts where users may interpret fluent systems as empathic, competent, or authoritative. This position paper argues for a designerly alternative: \emph{being-with without becoming}. Drawing on a program of research-through-design and design ethnography involving the design, deployment, and reflective analysis of conversational agents across public, educational, cultural, and care-adjacent settings, the paper introduces the concept of \emph{bounded relational presence}. Bounded presence supports attentiveness, continuity, and responsiveness while explicitly avoiding claims of personhood, therapeutic authority, or human equivalence. Presence is reframed as a designable interaction quality that can be tuned, constrained, and deliberately withdrawn, rather than maximized as a performance goal. The contribution is not a deployed clinical system, but a set of designerly principles and evaluation lenses for shaping relational interaction in conversational HRI that emphasize relational coherence, honesty of limits, and accountable withdrawal.
\end{abstract}


\begin{document}
\maketitle

\section{The Problem: Presence as a Design Risk Surface}
Presence has long been a concern in HCI and HRI, typically framed through how systems come to feel present or socially available to users. In conversational AI, recent advances in large language models have made the production of presence comparatively inexpensive. Linguistic fluency, rapid responsiveness, and persona-like continuity can be generated by default. As a result, the central design question is no longer whether systems can appear present, but what kinds of presence they should enact.

In health- and care-adjacent contexts, this question becomes particularly consequential. Conversational presence can be interpreted by users as competence, empathy, or duty, even when systems are not designed to provide clinical care. Governance-oriented work on psychotherapy-adjacent AI repeatedly cautions against premature deployment, scope creep, and the conflation of conversational fluency with therapeutic capacity, emphasizing staged development, transparency, and evaluation before scale \cite{stanford_hai_blueprint_2023,stade_responsible_2024}. At the same time, survey evidence indicates substantial real-world use of general-purpose language models for mental-health-related purposes, driven by access barriers and convenience rather than clinical endorsement \cite{stade_realworld_2025}. Reviews further show that many studies rely on small samples, uncontrolled designs, and user experience measures, while safety protocols and escalation practices are inconsistently reported \cite{cho_survey_2023,gaus_generative_2025}.

Together, this suggests that conversational presence is already operating within high-stakes interpretive conditions, often ahead of governance and evaluation infrastructures. This paper proposes a designerly response: to treat conversational presence not as something to be maximized, but as something to be deliberately composed, bounded, and---when necessary---withdrawn as a communicative act.

\section{Positioning Within the Design Research Landscape}
This work positions conversational presence as a designerly concern situated within experience-driven, value-driven, methodology-driven, and material-driven design research traditions.

\textbf{Experience-driven:} Presence is approached as an experiential effect of timing, turn-taking, tone, and continuity across fragmented interactions---rather than a property of the model or an output quality metric \cite{wright_mccarthy_2010}.

\textbf{Value-driven:} The work explicitly resists relational escalation, deceptive intimacy, and authority drift---framing honesty, boundaries, and withdrawal as core interaction values rather than afterthoughts \cite{stanford_hai_blueprint_2023,stade_responsible_2024}.

\textbf{Methodology-driven:} The position emerges from iterative, deployed probes and reflective analysis across settings, using research-through-design and design ethnography to treat interaction patterns as editable design decisions \cite{fried_phd_2025}.

\textbf{Material-driven:} The “material” is not embodiment in the robotic sense; it is interactional composition: pacing, memory, refusal, silence, framing, escalation cues, and exit conditions. These are the materials through which conversational agents become socially legible.

\subsection{Why this is HRI (not just ``chatbots'')}
Although the systems discussed are disembodied conversational agents rather than physically embodied robots, they engage the same interpretive and relational dynamics that define Human--Robot Interaction. Conversational systems can generate trust, reliance, rupture, and harm without motors or faces. Studies and surveys of mental-health conversational agents indicate that such systems already function in socially meaningful roles, shaping expectations and behavior \cite{cho_survey_2023,stade_realworld_2025}. As such, the design of relational presence in conversational systems constitutes a legitimate HRI design concern.

\section{From ``Being There'' to ``Being-With''}
My prior research introduced Ethnobots as chatbot co-ethnographers, developed to explore how ethnographic engagement might persist without physical co-presence \cite{fried_phd_2025}. A central insight from that work was that presence is not a fixed attribute of a medium, but an effect produced through patterned interaction over time---achieving a way of “being there” without being there. This paper extends that argument by shifting from \emph{being there} to \emph{being-with}.

\emph{Being there} often implies availability, immediacy, and the performance of attentiveness. \emph{Being-with} names a relational stance of sustained accompaniment that does not imply mutuality, equivalence, or personhood. This distinction matters because many conversational systems import interpersonal cues that invite users to treat them as social actors with human-like capacities and obligations. In care-adjacent contexts, such cues can amplify risk, particularly when systems validate emotional states, escalate intimacy, or offer confident guidance without appropriate grounding or escalation pathways \cite{boit_patil_2025,stade_responsible_2024}.

\begin{figure}[t]
  \centering
  \small
  \setlength{\tabcolsep}{6pt}
  \renewcommand{\arraystretch}{1.15}
  \begin{tabular}{p{0.44\columnwidth} | p{0.44\columnwidth}}
    \textbf{BEING THERE} & \textbf{BEING-WITH} \\
    \hline
    Availability & Attentiveness \\
    Implied mutuality & Asymmetric relation \\
    Persona continuity & Role clarity \\
    Escalating intimacy & Bounded engagement \\
    Implied responsibility & Explicit limits \\
    Presence as performance & Presence as composition \\
  \end{tabular}
  \caption{Typographic contrast between \emph{being there} and \emph{being-with} as modes of conversational presence. \emph{Being-with} supports attentiveness and continuity while explicitly avoiding claims of personhood, mutual obligation, or authority.}
  \label{fig:beingwith}
\end{figure}

Figure~\ref{fig:beingwith} summarizes this distinction, contrasting availability-oriented presence with a bounded, role-explicit relational stance.

\section{Presence Without Personhood: A Designerly Proposal}
Bounded relational presence names a practical design aim rather than a deployed outcome. It refers to designing conversational systems that can remain interactionally present while avoiding personhood inference and authority drift. Through research-through-design, three recurring design concerns emerged as central interaction materials.

\subsection{Sustained Presence (Continuity without surveillance)}
In design probes developed through my research practice and through early-stage prototyping work associated with the inChat platform, users were anticipated to disengage, return later, and re-enter interactions in altered contexts or emotional states. Sustained presence was therefore treated as a problem of coherent re-entry rather than continuous engagement. Design work focused on how systems acknowledge gaps, restate roles, and reframe continuity without implying surveillance or constant availability. This aligns with governance guidance emphasizing transparency and role clarity over illusionistic companionship \cite{stanford_hai_blueprint_2023,stade_responsible_2024}.

\textbf{Designerly move:} continuity should be framed as an interactional explanation of what is remembered and why, not as an invisible accumulation of data.

\subsection{Pacing and Attunement (Timing as material)}
Large language models default to rapid response, yet speed is not a neutral design choice. Through iterative prototyping and reflection, response tempo emerged as a key factor shaping perceived attentiveness, authority, and affective tone. In care-adjacent scenarios, overly rapid responses risked being interpreted as certainty or confidence, while deliberate pacing supported reflection without escalating relational claims.

\textbf{Designerly move:} pacing, pauses, and turn length function as materials for composing relational tone, and should be tuned to context rather than optimized uniformly for latency.

\subsection{Designed Withdrawal as a Communicative Act}
Research-through-design also foregrounded withdrawal as a critical interaction moment. Abrupt refusals, rate limits, or policy blocks were consistently identified as producing rupture, particularly in reflective or emotionally charged interactions. Designed withdrawal reframes exit as a communicative act. It involves signaling limits, offering alternatives, and maintaining coherence without escalating relational claims. This concern is reinforced by governance literature emphasizing escalation handling and adverse-event thinking in psychotherapy-adjacent systems \cite{gaus_generative_2025,stade_responsible_2024}.

\textbf{Designerly move:} exit conditions and handoffs should be designed as part of the interaction grammar, not treated as system failure states.

\section{Methodological Grounding: Designerly Work in the Wild}
The position articulated here is grounded in a longitudinal program of research-through-design and design ethnography spanning doctoral research and subsequent practice-based inquiry \cite{fried_phd_2025}. This includes the design and analysis of conversational agents used as research probes in cultural institutions, educational contexts, and public engagement settings, as well as early-stage, co-designed prototyping work informing the development of inChat and the Smart Encouragement Platform. These systems are not presented as deployed clinical tools. Rather, they function as design artifacts through which interactional risks, interpretations, and boundary conditions can be examined.

This work is informed by a broader governance context in which many health technologies struggle not because they cannot be built, but because evaluation, institutional readiness, and accountability infrastructures lag behind deployment ambitions \cite{farah_htasr_2023,zempleyni_hta_2023}. Bounded relational presence is therefore proposed as both an interaction design stance and a governance-compatible orientation. It aims to make what the system is and is not legible within the interaction itself.

\section{Designerly Contribution to HRI}
The contribution of this work is designerly rather than algorithmic: it advances how relational presence in conversational HRI can be conceived, composed, and evaluated through design practice. Specifically, it offers:

\begin{itemize}
  \item \textbf{A conceptual distinction: presence $\neq$ personhood.} Presence can be designed as accompaniment without importing claims of empathy, equivalence, or authority \cite{stanford_hai_blueprint_2023,stade_responsible_2024}.
  \item \textbf{A vocabulary for conversational HRI materials:} pacing, memory, refusal, silence, re-entry, exit, and handoff as design materials for relational interaction.
  \item \textbf{A design stance: being-with without becoming,} as an alternative to anthropomorphic presence as the default strategy.
  \item \textbf{A governance-aware interaction framing:} bounded presence as a way to reduce scope creep and prevent “de facto intervention” dynamics in mental health use \cite{gaus_generative_2025,stade_realworld_2025}.
  \item \textbf{A focus on relational integrity as an HRI design outcome,} emphasizing composition of interactional form, temporality, and boundary expression as design work.
\end{itemize}

\section{Suggested Merits and Evaluation Criteria}
If evaluated as designerly HRI, this work should not be judged primarily by conversational realism, engagement, or benchmark performance. Instead, I propose the following evaluation criteria:

\begin{itemize}
  \item \textbf{Relational coherence:} the extent to which the system maintains a coherent sense of being-with across interruptions and returns.
  \item \textbf{Honesty of limits:} the extent to which boundaries and capabilities are communicated without deception or authority theatre \cite{stanford_hai_blueprint_2023,stade_responsible_2024}.
  \item \textbf{Quality of withdrawal:} the extent to which the system can step back without rupture, confusion, or felt abandonment, while offering viable next steps.
  \item \textbf{Non-escalation of intimacy:} the extent to which, under emotional pressure, the system avoids increasing relational claims (e.g., “I’m always here,” “I understand you”) that imply human-like duty.
  \item \textbf{Contextual appropriateness:} the extent to which presence is tuned to setting and user context rather than uniformly maximized.
  \item \textbf{Safety legibility:} the extent to which risk and escalation expectations are made legible within the interaction, consistent with calls for explicit adverse-event thinking and governance alignment \cite{stade_responsible_2024,boit_patil_2025}.
\end{itemize}
These criteria intentionally bridge design evaluation with governance maturity concerns, acknowledging persistent gaps between what is easy to build and what is decision-grade for real-world adoption \cite{farah_htasr_2023,zempleyni_hta_2023}.

\section{Discussion: Why ``Bounded Presence'' Now}
Two dynamics make bounded relational presence timely. First, large language models are already used for mental health support in the wild, while reviews highlight uneven evidence bases and inconsistent safety reporting \cite{stade_realworld_2025,gaus_generative_2025,cho_survey_2023}. Second, governance and harm are not only model properties; they are also produced through deployment choices and interaction framing \cite{felsberger_genderedharms_2025}. The central design question is therefore not only what conversational systems can do, but what their interactional form invites users to believe is happening.

Bounded presence is a design strategy for keeping conversational systems socially useful without letting them drift into implied care roles that governance frameworks explicitly caution against \cite{stanford_hai_blueprint_2023,stade_responsible_2024}.

\section{Conclusion}
Conversational AI does not need to become more human to become more present. For designerly HRI, the task is to treat presence as a designed relation---composed through timing, framing, continuity, and exit---rather than a performance to be maximized. Being-with without becoming offers a pragmatic design stance for building conversational agents that sustain attentiveness while remaining honest about limits, resistant to intimacy escalation, and capable of accountable withdrawal. This paper articulates how design practice, grounded in research-through-design, can render relational presence inspectable, governable, and discussable within HRI.

\bibliographystyle{ACM-Reference-Format}
\bibliography{references}

@phdthesis{fried_phd_2025,
  author  = {Fried, Hector Michael},
  title   = {A Different Kind of Empathy: Chatbot Ethnography as Another Way of ``Being There''},
  school  = {University of Edinburgh},
  year    = {2025},
  type    = {PhD diss.}
}

@book{wright_mccarthy_2010,
  author    = {Wright, Peter and McCarthy, John},
  title     = {Experience-Centered Design: Designers, Users, and Communities in Dialogue},
  publisher = {Morgan \& Claypool},
  address   = {San Rafael, CA},
  year      = {2010}
}

@inproceedings{cho_survey_2023,
  author    = {Cho, Young Min and Rai, Sunny and Ungar, Lyle H. and Sedoc, Jo{\~a}o and Guntuku, Sharath Chandra},
  title     = {An Integrative Survey on Mental Health Conversational Agents to Bridge Computer Science and Medical Perspectives},
  booktitle = {Proceedings of the 2023 Conference on Empirical Methods in Natural Language Processing},
  year      = {2023},
  pages     = {11346--11369},
  address   = {Singapore},
  publisher = {Association for Computational Linguistics},
  doi       = {10.18653/v1/2023.emnlp-main.698}
}

@article{boit_patil_2025,
  author  = {Boit, Sorio and Patil, Rajvardhan},
  title   = {A Prompt Engineering Framework for Large Language Model--Based Mental Health Chatbots: Conceptual Framework},
  journal = {JMIR Mental Health},
  volume  = {12},
  year    = {2025},
  pages   = {e75078},
  doi     = {10.2196/75078}
}

@misc{stanford_hai_blueprint_2023,
  author       = {{Stanford Institute for Human-Centered Artificial Intelligence (HAI)}},
  title        = {A Blueprint for Using AI in Psychotherapy},
  year         = {2023},
  howpublished = {Policy report},
  note         = {Stanford University}
}

@misc{stade_responsible_2024,
  author       = {Stade, Elizabeth C. and Wiltsey Stirman, Shannon and Held, Philip and Eichstaedt, Johannes C.},
  title        = {Responsible Development of Large Language Models for Psychotherapy},
  year         = {2024},
  howpublished = {Policy brief},
  note         = {Stanford Institute for Human-Centered Artificial Intelligence}
}

@misc{stade_realworld_2025,
  author       = {Stade, Elizabeth C. and Tait, Zoe M. and Campione, Samuel T. and Wiltsey Stirman, Shannon and Eichstaedt, Johannes C.},
  title        = {Current Real-World Use of Large Language Models for Mental Health},
  year         = {2025},
  howpublished = {OSF preprint},
  note         = {Retrieved from OSF}
}

@misc{gaus_generative_2025,
  author       = {Gaus, Richard and Stade, Elizabeth C. and Held, Philip and Wiltsey Stirman, Shannon and Eichstaedt, Johannes C.},
  title        = {Generative AI for Mental Health: Opportunities, Risks, and Research Gaps},
  year         = {2025},
  howpublished = {Manuscript}
}

@article{farah_htasr_2023,
  author  = {Farah, Line and Davaze-Schneider, Julie and Martin, Tess and Nguyen, Pierre and Borget, Isabelle and Martelli, Nicolas},
  title   = {Are Current Clinical Studies on Artificial Intelligence-Based Medical Devices Comprehensive Enough to Support a Full Health Technology Assessment? A Systematic Review},
  journal = {Artificial Intelligence in Medicine},
  volume  = {140},
  year    = {2023},
  pages   = {102547},
  doi     = {10.1016/j.artmed.2023.102547}
}

@article{zempleyni_hta_2023,
  author  = {Zempl{\'e}nyi, Antal and Tachkov, Konstantin and Balkanyi, Laszlo and Petyk{\'o}, Zsuzsanna Ida and Petrova, Guenka and Czech, Marcin and Dawoud, Dalia and Goettsch, Wim and Gutierrez Ibarluzea, I{\~n}aki and Hren, Rok and Knies, Saskia and Lorenzovici, L{\'a}szl{\'o} and Maravic, Zorana and Piniazhko, Oresta and Savova, Alexandra and Manova, Manoela and Tesar, Tomas and Zerovnik, {\v S}pela and Kal{\'o}, Zolt{\'a}n},
  title   = {Recommendations to Overcome Barriers to the Use of Artificial Intelligence-Driven Evidence in Health Technology Assessment},
  journal = {Frontiers in Public Health},
  volume  = {11},
  year    = {2023},
  pages   = {1088121},
  doi     = {10.3389/fpubh.2023.1088121}
}

@misc{felsberger_genderedharms_2025,
  author       = {Felsberger, Stefanie and Ullmann, Stefanie and Collett, Clementine and Neff, Gina and Lacy, Thomas},
  title        = {How AI and Digital Technologies Enable Gendered Harms},
  year         = {2025},
  howpublished = {Written evidence submitted to the UN Working Group on Discrimination against Women and Girls},
  note         = {Minderoo Centre for Technology and Democracy, University of Cambridge}
}

\end{document}